\documentclass[]{spie}

\usepackage{amsmath,amsfonts,amssymb}
\usepackage{graphicx}
\graphicspath{{.}} 
\usepackage{subcaption}
\usepackage{sidecap}
\usepackage[colorlinks=true, allcolors=blue]{hyperref}

\title{Panoramic optical and near-infrared SETI instrument: prototype design and testing}

\author[a,b]{Maren Cosens}
\author[a]{J\'er\^ome Maire}
\author[a,b]{Shelley A. Wright}
\author[c]{Franklin Antonio}
\author[d]{Michael Aronson}
\author[e]{Samuel A. Chaim-Weismann}
\author[f]{Frank D. Drake}
\author[g]{Paul Horowitz}
\author[h]{Andrew W. Howard}
\author[i]{Rick Raffanti}
\author[e,f,j,k]{Andrew P.V. Siemion}
\author[l]{Remington P.S. Stone}
\author[m]{Richard R. Treffers}
\author[n]{Avinash Uttamchandani}
\author[e,o]{Dan Werthimer}
\affil[a]{Center for Astrophysics \& Space Sciences, University of California San Diego, USA}
\affil[b]{Department of Physics, University of California San Diego, USA}
\affil[c]{Qualcomm Inc., San Diego, USA}
\affil[d]{Electronics Packaging Man, San Diego, USA}
\affil[e]{Department of Astronomy, University of California Berkeley, USA}
\affil[f]{SETI Institute, Mountain View, USA}
\affil[g]{Department of Physics, Harvard University, USA}
\affil[h]{Astronomy Department, California Institute of Technology, USA}
\affil[i]{Techne Instruments, Oakland, USA}
\affil[j]{Radboud University, Nijmegen, Netherlands}
\affil[k]{Institute of Space Sciences and Astronomy, University of Malta}
\affil[l]{University of California Observatories, Lick Observatory, USA}
\affil[m]{Starman Systems, Alamo, USA}
\affil[n]{Nonholonomy, LLC, Cambridge, USA}
\affil[o]{Space Sciences Laboratory, University of California Berkeley, USA}

\authorinfo{Send correspondence to Maren Cosens \\E-mail: mcosens@ucsd.edu}

\begin{document} 
\maketitle

\begin{abstract}
The Pulsed All-sky Near-infrared Optical Search for ExtraTerrestrial Intelligence (PANOSETI) is an instrument program that aims to search for fast transient signals (nano-second to seconds) of artificial or astrophysical origin. The PANOSETI instrument objective is to sample the entire observable sky during all observable time at optical and near-infrared wavelengths over 300 - 1650 nm \cite{Wright2018}. The PANOSETI instrument is designed with a number of modular telescope units using Fresnel lenses ($\sim$0.5m) arranged on two geodesic domes in order to maximize sky coverage \cite{Maire2018}. We present the prototype design and tests of these modular Fresnel telescope units. This consists of the design of mechanical components such as the lens mounting and module frame. One of the most important goals of the modules is to maintain the characteristics of the Fresnel lens under a variety of operating conditions. We discuss how we account for a range of operating temperatures, humidity, and module orientations in our design in order to minimize undesirable changes to our focal length or angular resolution.
\end{abstract}

\keywords{Search for ExtraTerrestrial Intelligence, SETI, Techno-signatures, Astrophysical transients, Fresnel lenses, Fresnel telescopes, All-sky}

\section{INTRODUCTION}\label{sec:intro}
The PANOSETI instrument is designed to sample the entire observable sky during all observable time at optical and near-infrared wavelengths in search of pulsed signals of astrophysical or artificial origin. Signals of artificial origin may come from pulsed laser communication \cite{Horowitz1993, Howard2004} or leakage from energy transmission (e.g., to propel spacecraft with light sails \cite{Guillochon2015}). These types of pulses could be easily detected with a 10m class telescope, being $\sim10^4\times$ brighter than a Sun-like star \cite{Howard2004}. Previous optical SETI searches have either been targeted or wide field surveys making use of only a single aperture \cite{Howard2004, Wright2014}. This leads to low dwell times per source, greatly reducing the chances of detecting intermittent signals.

\begin{figure}[h]
\begin{subfigure}{0.49\textwidth}
\centering
\includegraphics[scale=0.33]{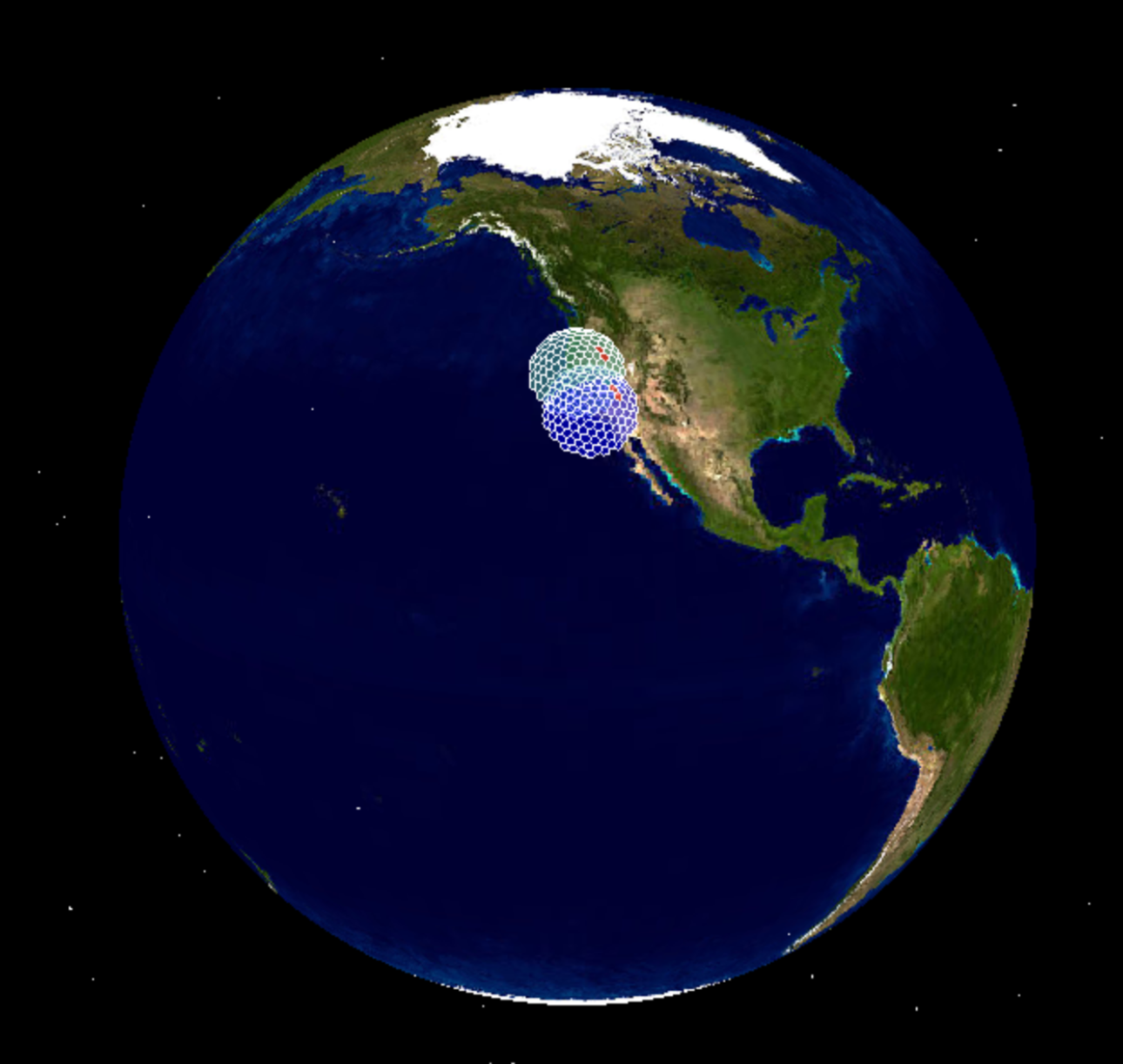} 
\caption{}
\end{subfigure}
\begin{subfigure}{0.49\textwidth}
\centering
\includegraphics[scale=0.47]{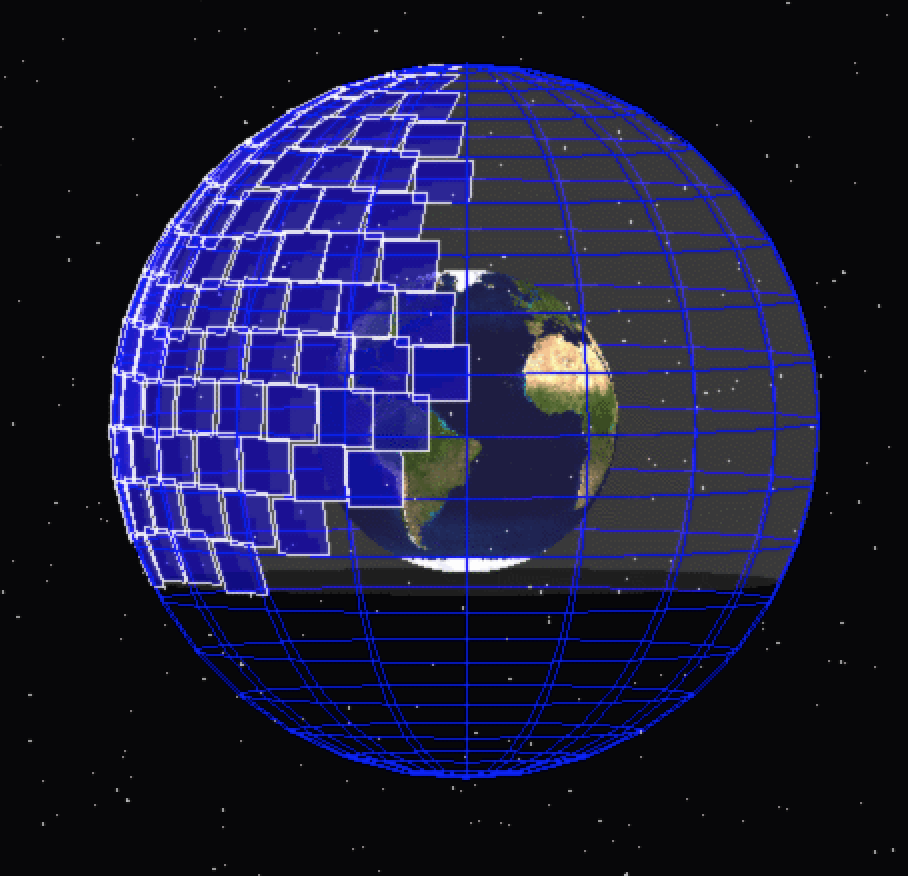}
\caption{}
\end{subfigure}
\caption{(a): Two PANOSETI geodesic dome facilities planned to be located at Lick Observatory near San Jose, CA and Mt.~Laguna Observatory near San Diego, CA ($\sim$690km or $\sim$430 miles apart). Each facility will have some one-hundred individual telescope modules in the final phase of the PANOSETI design. Both locations will observe the same field of view to confirm events via coincidence detection. The blue hexagons correspond to modules operating at optical wavelengths, while the red hexagons represent the two modules at each site operating at the near-infrared. (b): Instantaneous field of view from the two dome facilities \cite{Maire2018}. Each square represents the projection of a 32x32 pixel Hamamatsu silicon photo-multiplier (SiPM) detector array in an individual telescope module containing a $\sim$0.5m Fresnel lens aperture.}
\label{fig:FoV}
\end{figure}

PANOSETI aims to be the first dedicated all-sky optical and near-infrared SETI experiment targeting the entire northern hemisphere with $\gtrsim$2 steradians of instantaneous sky coverage \cite{Maire2018}. Geodesic domes at two sites--Mt.~Laguna and Lick Observatory--will be used for coincidence detection of signals with each dome containing some one-hundred individual telescope modules. Figure \ref{fig:FoV} shows an illustration of the sky coverage from these two observing facilities.

\begin{figure}[h]
\begin{subfigure}{0.49\textwidth}
\centering
\includegraphics[scale=0.15]{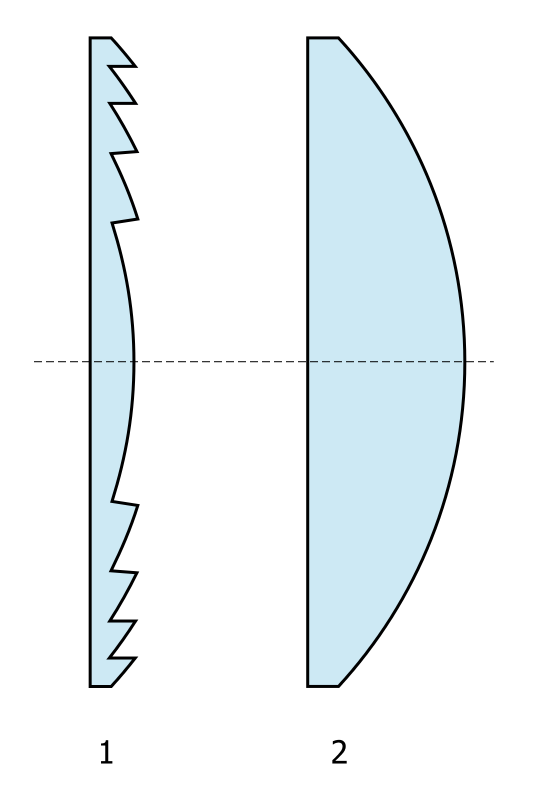} 
\caption{}
\end{subfigure}
\begin{subfigure}{0.49\textwidth}
\centering
\includegraphics[scale=0.5]{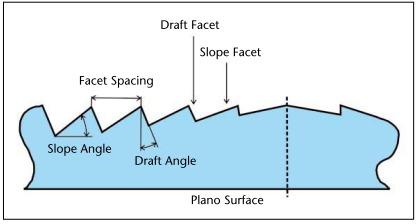}
\caption{}
\end{subfigure}
\caption{(a): Comparison of traditional convex lens and a Fresnel lens of equivalent surface curvature. Significantly less material volume is required for the Fresnel lens. (b): Image credit: Davis \& K\"{u}hnlenz (2011)\cite{Davis2011}. Illustration of Fresnel lens schematic showing the terminology used to describe the lens as well as the potential for losses in the shadow of the draft facet.}
\label{fig:lens_schem}
\end{figure}

Each telescope module will make use of a $\sim$0.5m f/1 Fresnel lens as a lightweight, cost-effective alternative to traditional apertures, ideal for use in a large scale instrument with a total of some two-hundred apertures between the two domes. A Fresnel lens makes use of concentric grooves that maintain the curvature of a traditional lens in a thinner, lighter format \cite{Davis2011} (see Figure \ref{fig:lens_schem} for schematic of Fresnel lens characteristics and terminology). These modules will be self-contained and interchangeable to be placed at any location in the geodesic dome. Each module will be fixed to the dome and rotated to the proper position angle (PA) to maximize sky coverage \cite{Maire2018}.

The individual modules required for the full PANOSETI instrument will be developed in phases detailed in Section \ref{sec:phases}. The primary mechanical loads impacting these modules are discussed in Section \ref{sec:loads} and the key components of each module design are discussed in Section \ref{sec:module}. We summarize the component design for each module phase in Section \ref{sec:summary}.

\section{Prototype Phases}\label{sec:phases}
The PANOSETI project is broken up into a number of testing phases with specific goals for each phase. The individual aperture modules used in each phase will be described here briefly with the mechanical design detailed further in Section \ref{sec:module}. The final modules will be secured in a geodesic dome from two observatory sites.

\subsection{Beta-1}
The initial Beta-1 module was designed as a proof of concept using a single Fresnel lens, making use of an Edmund Optics 46-392 lens (referred to here as the Edmund Optics lens for simplicity). It was used to test the opto-mechanical designs, focus and image quality at the detector, and sky background levels. This module was designed to be cost effective and easily assembled due to the large number of modules required for the final instrument design. A key focus was on designing a secure lens mounting and easily removable baffling opaque at both visible and near-infrared wavelengths.

The Beta-1 module is being used for ongoing measurements of the sky background level at Mt.~Laguna Observatory under a variety of conditions. These measurements will inform the design of electronics and software to be used with the final instrument.

\begin{SCfigure}
\includegraphics[width=0.25\textwidth]{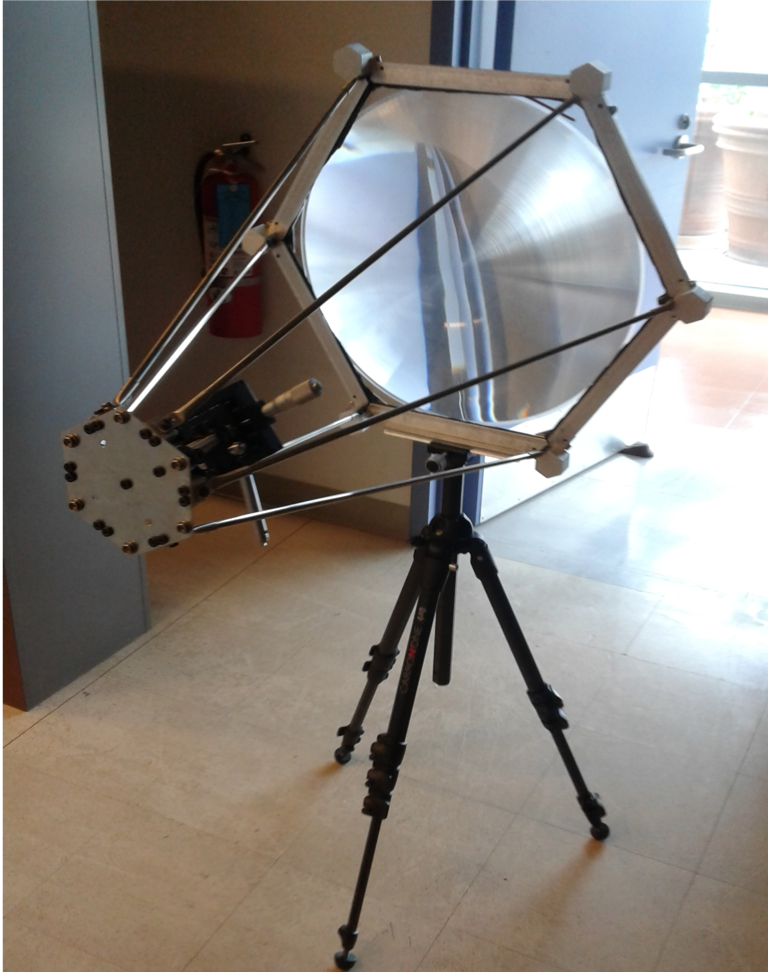} 
\caption{Assembled Beta-1 module utilizing Edmund Optics 46-392 lens with opaque baffling removed. Lens frame is made up of 6 pieces to secure the Fresnel lens and acrylic protector on the outside of the module. A neoprene baffling (not pictured) is wrapped around the module and secured by snaps to be easily removable.}
\label{fig:beta1}
\end{SCfigure}

\subsection{Prototype Module}
The prototype module was designed to test specific components to be improved from Beta-1 and incorporated into the design of modules for the final geodesic dome. These primarily consist of the lens mounting, addition of a focus stage, and incorporation of custom electronics in what is referred to as the quadrant board. Four of these modules will be produced, with two placed at each observatory location for preliminary testing and science operations. This will allow for further testing of the opto-mechanical design as well as the first operational test of the custom electronics and software. In contrast to the Beta-1 module, the Prototype module (and modules going forward) will make use of an Orafol SC214 Fresnel lens (referred to here as the Orafol lens).

\subsection{Future Modules}
\begin{SCfigure}
\includegraphics[width=0.35\textwidth]{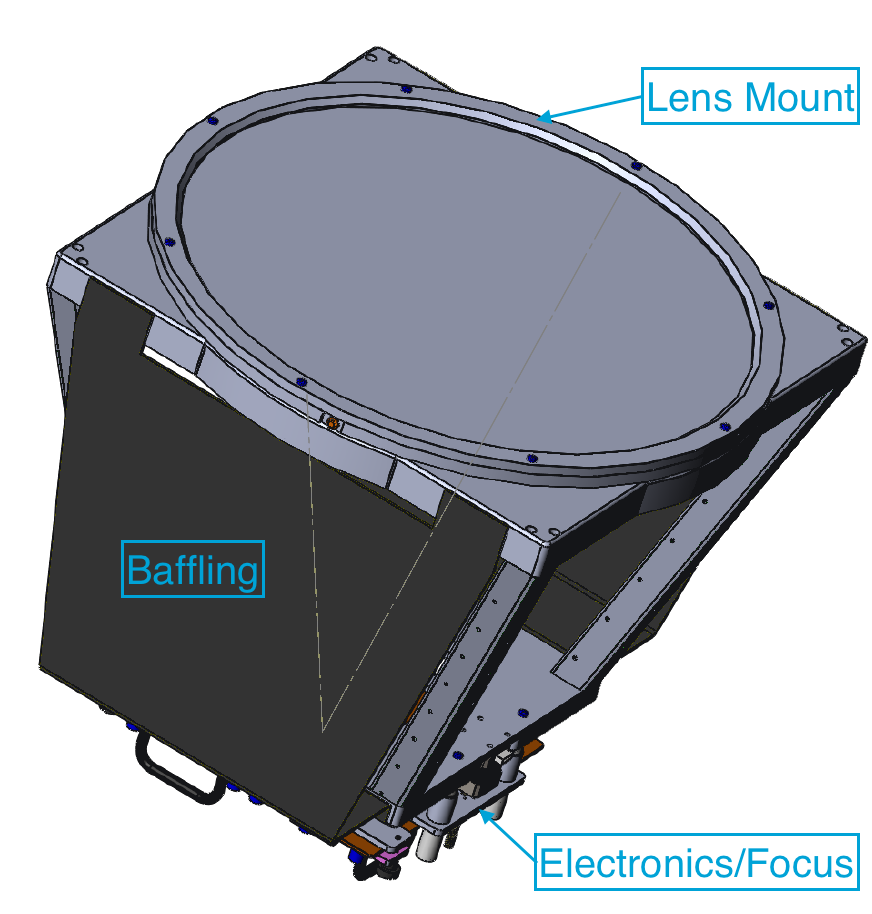} 
\caption{Design of final module to be placed in geodesic dome structure tiling the sky. Lens mounting scheme from the prototype module will be incorporated into an all aluminum frame (gray) with either aluminum panels (black) or neoprene used as baffling. The aluminum frame will allow for straightforward manufacturing and interfacing with the geodesic dome. A custom focus stage will be mounted with electronics on the back end of the module to adjust for changing observing conditions and variations in the lens focal length.}
\label{fig:final_module}
\end{SCfigure}

The final module will incorporate components of the Prototype module like the lens mounting and focus stage with any improvements found necessary during testing. Most significantly the electronics will consist of four boards like the quadrant board used in the previous module, providing four times the detector area. A mock-up of this module is shown in Figure \ref{fig:final_module}, and will be mounted to each of the openings in the geodesic domes and rotated to the desired PA. These modules will be identical and interchangeable to any dome location.

\section{Expected Loads}\label{sec:loads}
We expect a number of forces acting on each module and Fresnel lens in particular which, if not properly accounted for, can greatly impact the performance of the telescope. Facet spacing as well as slope and draft angles are key characteristics of Fresnel lenses that are affected by any deformation or deflection of the lens, and even small deflections or deformations can have a significant impact on the focal length and angular resolution \cite{Maire2018}. We discuss here our calculations of the expected load and deformation from gravity, wind, and thermal expansion/contraction of the lens. Evaluation of these loads are of even greater importance than a traditional telescope as there will be no enclosure adding protection from the elements during operation of the instrument.

\subsection{Gravity}\label{sec:weight}
As these Fresnel modules will tessellate a geodesic dome, the orientation will vary slightly from module to module. Thus the direction of the gravitational force on each lens (weight) will vary with module position. Since each module should be interchangeable with any other we design for the worst case scenario for each load. In the case of lens weight, the main concern is deflection of the lens which will be most pronounced in the module at zenith where the lens is parallel to the ground.

To investigate this deformation we model the lens as a beam with a rectangular cross section supported at two points. This gives a moment of inertia described by:
\begin{equation}
    I = \frac{1}{12} b h^3 = \frac{1}{12} d_0 t_0^3
\end{equation}
with $d_0$ and $t_0$ representing the nominal lens diameter and thickness respectively (without deformation).
We then calculate the expected deflection at the center of the lens:

\begin{eqnarray}
\begin{array}{cc}
    \delta = (\frac{W x}{24 E I})(L^3 - 2 L x^2 + x^3) \\
    \delta_{max} (x=\frac{L}{2}=\frac{d_0}{2}) = \frac{5 W d_0^3}{384 E I} \\
\end{array}
\end{eqnarray}
$W$ is the weight of the lens, and $E$ is Young's modulus for the acrylic Fresnel lens (approximated as $3.2\times10^9 N \ m^{-2}$).

For the Edmund Optics lens used in the Beta-1 module this gives an expected deflection of 2.3 mm at the center and 6.2 mm deflection for the Orafol lens planned to be used in the Prototype and future modules. This deflection would cause too great a change in focal length and distortion of the image, so a stiffener must be included in the design (see Section \ref{sec:lens_mount}).

\subsection{Thermal Expansion}\label{sec:thermal}
The nature of the populated geodesic dome design makes temperature regulation challenging, and the modules will be exposed to a wide range of operating temperatures. The lenses on the outside of the dome will largely be exposed to ambient conditions which can vary significantly throughout the year at Lick Observatory and Mt.~Laguna Observatory. At night we plan for operating temperatures to vary between $-$20$^\circ$C to 40$^\circ$C (0$^\circ$F to 100$^\circ$F) throughout the year\footnote{We use an overestimate of the maximum operating temperature based on current records, but this provides a factor of safety in our designs that is particularly desirable given the variation in acrylic material properties.}.

In addition to the wide range of operating temperatures, this instrument is unique in that the lenses are not made of glass but acrylic. This results in the lenses being $\sim$10$\times$ more prone to deformation with changes in temperature. As the exact coefficient of thermal expansion will vary by lens manufacture (due to material or manufacturing differences) and is not often specified, we use a typical value of $\alpha = 75\times10^{-6} \rm \ m \ m^{-1} \ K^{-1}$ for both the Edmund optics lens and the Orafol lens. Expansion and contraction of the lens is described by the equation:

\begin{equation}
    \Delta d = \alpha  d_0 \Delta T
\end{equation}
with $\Delta d$ giving the expected change in lens diameter. This was also used for the thickness of the lens (a much smaller effect).

For both lenses this resulted in a change in diameter of +0.55mm between room temperature and 40$^\circ$C, and a change of $-$1.4\,mm between room temperature and $-$20$^\circ$C. Changes in lens diameter also cause proportional changes to the facet spacing and thus the focal length which must be accounted for in the focus of the detector.

An increase in lens diameter also results in a force on the frame if it is held fixed on opposite sides given by:

\begin{equation}
    F_t =\frac{\Delta d}{d_0} A E
\end{equation}
with $A$ being the area of fixed contact between the lens and mounting. This results in a force of $\sim$2.8kN for the Edmund Optics lens and $\sim$1.8\,kN for the Orafol lens with 1/6 of the lens edge used as the contact area (based on Beta-1 module design; see Section \ref{sec:lens_mount}). If this force is large enough then not only will stresses cause distortion of the lens and grooves, but it could cause the lens to bow. To investigate this we model the lens as a column and use Euler's formula to find the critical load for column buckling:
\begin{equation}
    P_{cr} = \frac{\pi^2 E I}{L_{eff}^2}
\end{equation}
where the effective length, $L_{eff}$, is dependent on the end conditions. For two fixed ends $L_{eff}=0.65 d_0$, but allowing one side to expand freely gives $L_{eff}=2.1 d_0$. Comparing $P_{cr}$ with the force experienced in each mounting configuration we find which configurations would lead to buckling. If both ends of the lens are held in a fixed position the critical load is $\sim$350\,N, much less than the force due to thermal expansion. Therefore, free expansion of the lens must be allowed in the lens mounting to prevent bowing of the lens as the temperature increases.

\subsection{Wind}\label{sec:wind}
Like for gravitational loads, the impact of wind will vary by module location. For the worst case lens deflection due to wind load we investigate a module with the lens directly perpendicular to an $\rm80 \ km \ h^{-1}$ (50mph) wind. The force due to wind is:
\begin{equation}
    F_w = \frac{1}{2} c_D \rho v^2 A
\end{equation}
Using the density of air at Standard Temperature and Pressure ($\rho = 1.225 \rm \ kg \ m^{-3}$), the drag coefficient for a plate perpendicular to the wind ($c_D = 1.17$), and wind speed of $\rm80 \ km \ h^{-1}$ ($v=22.2 \rm \ m \ s^{-1}$), we find an applied force of 59.8\,N on a surface the size of the Orafol lens and 61.9 N for the area of the Edmund optics lens. This would cause greater than a 10\,cm deflection at the center, further indicating the necessity of a stiffener beam.

\section{Module Components} \label{sec:module}
Each telescope module has a number of important components which must be designed to function together and to account for the loads described in the previous section. The primary components of these modules are the Fresnel lens itself, the lens mount, the frame, the focus stage with electronics, and the baffling.

\subsection{Fresnel Lens}\label{sec:lens}
A Fresnel lens is a unique optical element that provides a number of advantages for use in the PANOSETI telescope modules (with of course some trade-offs). Fresnel lenses are often made from injection or compression molded acrylic to produce cost-effective, lightweight lenses. The main disadvantage of a Fresnel lens is transmission losses on the draft facet (see Figure \ref{fig:lens_schem}) where the tip of one groove essentially causes a shadow on the adjacent groove. This can be mitigated by reducing the draft angle--producing a steeper drop between grooves--but this cannot be reduced to zero due to manufacturing constraints.

Even with the losses due to a non-zero draft angle, the transmission of Fresnel lenses is still high. At an f-number (the ratio of focal length to lens diameter) of 1.0, the theoretical transmission of a concentrating Fresnel lens is greater than 80\% \cite{Davis2011}.

To find the best lens for use in the PANOSETI modules, a number of lenses from different manufacturers were characterized \cite{Maire2018}. In the Beta-1 module the Edmund Optics 46-392 lens is used, with a diameter of 470mm and a focal length of 457 mm, giving an f-number just under 1. For the prototype and future modules, the Orafol SC214 lens is planned to be used. This lens has a diameter of 461mm and a focal length of 608mm (f-number$\sim$1.3). The Orafol SC214 was chosen due to decreased spot size and improved angular resolution compared to the Edmund Optics 46-392 \cite{Maire2018}. Characteristic properties of these two lenses are given in Table \ref{tbl:properties} along with the effect of loads discussed in Section \ref{sec:loads}.

\subsection{Lens Mount}\label{sec:lens_mount}
Due to the wide range of temperatures at each observatory location, forces from thermal expansion could damage the lens or lead to bowing if the mount is too rigid. As discussed in Section \ref{sec:thermal}, we find that keeping the lens fixed on all sides will cause thermal expansion forces to exceed a critical value and cause deflection of the lens. 

\begin{figure}[h]
\begin{subfigure}{0.49\textwidth}
\centering
\includegraphics[scale=0.5]{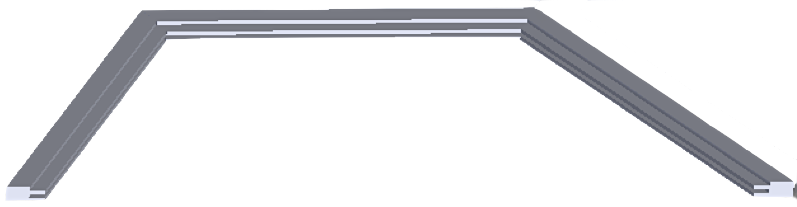} 
\caption{}
\end{subfigure}
\begin{subfigure}{0.49\textwidth}
\centering
\includegraphics[scale=0.5]{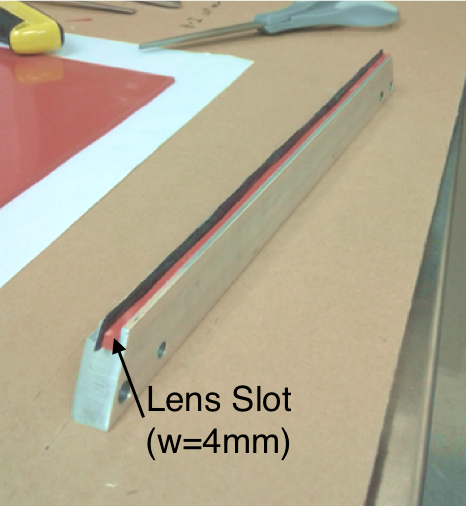}
\caption{}
\end{subfigure}
\caption{(a): Sketch of half of the Beta-1 lens frame showing oversized slots allowing thermal expansion and contraction of the Fresnel lens. Top step for a protective acrylic plate secured by an aluminum fastener plate at each corner of the frame. (b): Single piece of lens frame. 3\,mm thick silicone rubber (red) is placed at the bottom of the lens slot and 1\,mm thick neoprene (black) on the side to secure the lens without prohibiting expansion. The step for the protective acrylic is on the left in this image (outside of the module).}
\label{fig:beta_lensmount}
\end{figure}

In the Beta-1 module we sought to address this with a slotted lens mount design with extra space in the slots for the lens to expand. Since the lens cannot be free to move translationally within the lens mount (moving the center focus position relative to the detector) we used neoprene and silicone rubber (durometer: 50A) to fill the extra space. This rubber was stiff enough to not deform under the small weight of the lens, but soft enough to deform under the thermal expansion force from the lens. We used 3\,mm thick silicone rubber on the edge of the lens mount and 1\,mm thick neoprene rubber on the side to allow thermal expansion of the lens both radially and perpendicularly. Six pieces of machined aluminum make up the hexagonal Beta-1 lens mount, each with a slot for the rubber and lens as well as a step for a protective acrylic layer 1mm above the lens to prevent damaging the groove facets. The acrylic is cut into a hexagonal shape to mount easily in the frame. Gaps between the circular Fresnel lens and each corner of the hexagonal lens mount allow for airflow between the lens and the protective acrylic, preventing moisture from condensing in this small gap. The assembled lens mount is shown with the module in Figure \ref{fig:beta1}, and Figure \ref{fig:beta_lensmount} shows the individual aluminum members of this mount in more detail. 

This slot and rubber design works well for fixing the center position of the lens and absorbing expansion forces for a small temperature range (e.g., during a single night), but with large deviations from room temperature (for our expected operating temperature range) buckling of the lens can occur. The rubber is too hard to be readily deformable, leading to a radial force inward on the lens. For high temperatures this will result in bowing of the lens. For the next design iteration of the Prototype module we aim to address this problem with a new lens mount.

The Prototype module lens mount makes use of an aluminum ring of larger diameter than the lens with a step to support it and two set screws to fix the position. This allows for free expansion everywhere preventing stress on the acrylic Fresnel lens while also fixing the position of the lens and groove facets relative to the plane of the detector. An aluminum spacer ring will separate the Fresnel lens from the acrylic protector plate which will be secured via the same set screw method.

As this module is designed to be used at every dome location, there are a number of loads that were not present or addressed with the Beta-1 module. The first is the weight of the lens, causing sag at the center which is primarily an issue at or near zenith when the lens is parallel to the ground. With the fixed module positions in the final PANOSETI design, lens sag will be more noticeable than with Beta-1 which is stored pointing at the horizon and operated at changing elevation. The Edmund Optics lens in the Beta-1 module is also thicker (and therefore stiffer) than the Orafol lens to be used in the Prototype module and beyond. This leads to an increase in estimated sag at the lens center from 2.3\,mm to 6.2\,mm with the Orafol lens pointed at zenith and over 100\,mm of deflection for lenses perpendicular to a $\rm80 \ km \ h^{-1}$ wind. This large deflection causes distortion of the facet spacing and angle as well as the focal length for the Fresnel lens and changes to the angle of incidence for the acrylic protector plate. To prevent lens sag from degrading image quality a stiffening beam is added below both the acrylic protector plate (2.5\,mm$\times$10\,mm) and the Fresnel lens (4\,mm$\times$10\,mm). These stiffening beams reduce the lens sag due to gravity to $\sim$0.1\,mm and the deflection from $\rm80 \ km \ h^{-1}$ winds to $\sim$1.3\,mm. This deflection will not damage the lenses with large gusts of wind, and sag due to gravity will not limit data quality.

\begin{figure}[h]
\centering
\includegraphics[scale=0.5]{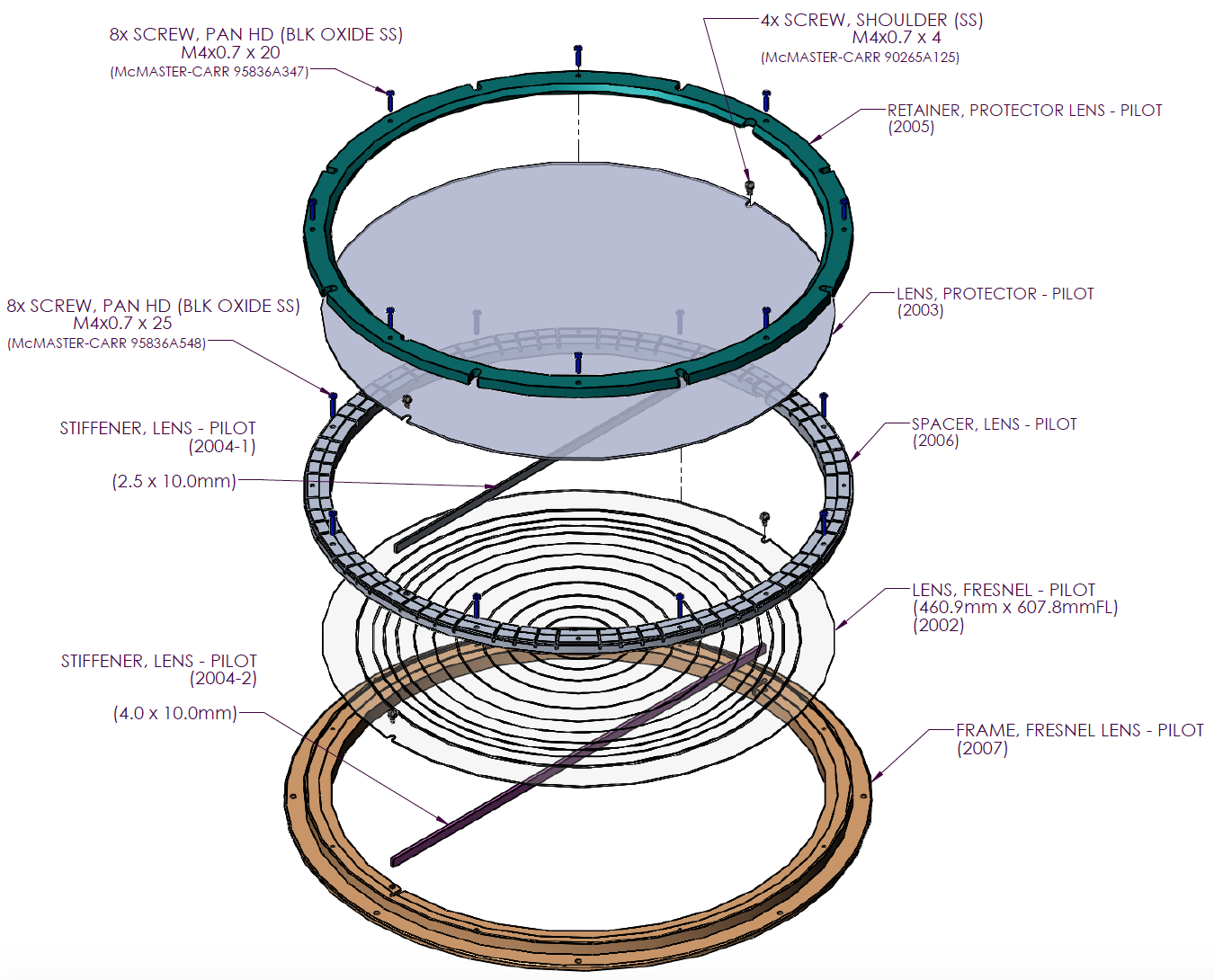} 
\caption{Exploded view of prototype lens mount including cross bar supports to prevent lens sag, ventilation holes to prevent condensation, and set screws to fix lens position but still allow free expansion with increases in temperature. All components are aluminum with textures applied to differentiate between parts.}
\label{fig:proto_lens}
\end{figure}

To mitigate the problem of condensation between the Fresnel lens and protective acrylic ventilation holes will be drilled in the aluminum spacer. A sketch of the Prototype lens mount is shown in Figure \ref{fig:proto_lens} detailing these components. This lens mount is planned to be incorporated into the final module design with minimal modifications as needed.

\subsection{Frame}\label{sec:frame}

\begin{figure}[h]
\centering
\includegraphics[width=.7\textwidth]{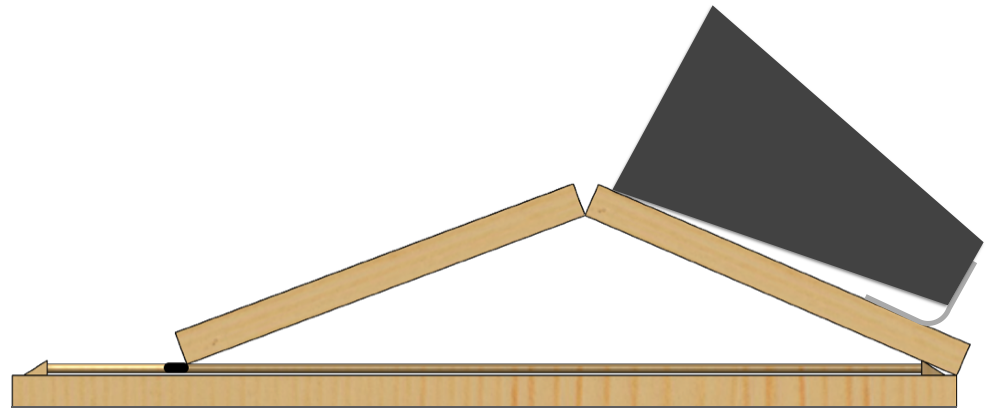} 
\caption{Illustration of the mounting used for observations with the Beta-1 telescope module. Hinges at all joints and a sleeve around the base guide rod allow for adjustments to elevation. The entire mount is rotated for further pointing adjustments. The module is secured to the base at the front and back via screws to allow for easy assembly and dis-assembly.}
\label{fig:beta_mount}
\end{figure}

The design of the module frame varies greatly between iterations due to the varying goals of each project phase. The Beta-1 module makes use of cost effective steel struts between the hexagonal lens mount and a hexagonal back-plate for mounting the detector or camera (depending on what test is being conducted). These struts are made from lightweight, hollow rods to make the module easily transportable. The frame extends past the focal plane of the lens to allow for focus adjustments and testing with a Ninox 640 InGaAs camera as well as measurements with a single Hamamatsu silicon photo-multiplier (SiPM) pixel planned for use in the full detector. This frame is attached to a wood mount for adjusting the pointing elevation for on-sky observations. This consists of a wood base and guide rod. Two additional pieces of wood attached with hinges form a triangle with a sliding joint along the guide rod to change the angle of the module (see Figure \ref{fig:beta_mount}). The module can be easily removed for transportation and testing in the lab.

A more rigid design is planned for the Prototype module, taking advantage of the single fixed operating location and electronics configuration with the frame no longer extending past the focal plane. The aluminum lens mount will attach to four aluminum struts which extend to a square back plate. A focus stage with electronics and detectors will be located behind this plate, requiring a square cutout in the plate to allow light to pass through onto the detector. This frame and electronics will all be contained within a wood box for baffling and mounting to a Dobsonian telescope mount for adjusting the elevation of the module pointing at each observing location. A door in the wood box will allow access to the module to make any adjustments to electronics.

Much of the frame design from the Prototype module will be adapted into the final module design. The same aluminum struts will connect the lens mount to an aluminum back plate on which the electronics and focus stage will be attached. The final module will not be contained within a wood box, owing to the required longevity of the final module and space constraints within the geodesic dome. Attachment points will be added to this frame (likely at the lens mount) to interface with the geodesic dome at an adjustable orientation.

\subsection{Back-end Electronics and Focus Stage}\label{sec:focus}
Custom designed electronics will be used with Multi-Pixel Photon Counting (MPPC) detectors. For the Beta-1 module tests either the camera or single pixel detector are mounted onto a Thorlabs optical post and manually adjusted to align with the lens center and focal plane. A manual positioning and focus is sufficient for these tests, but subsequent modules require more robust mounting and focus adjustment.

\begin{figure}[h]
\begin{subfigure}{0.49\textwidth}
\centering
\includegraphics[scale=0.3]{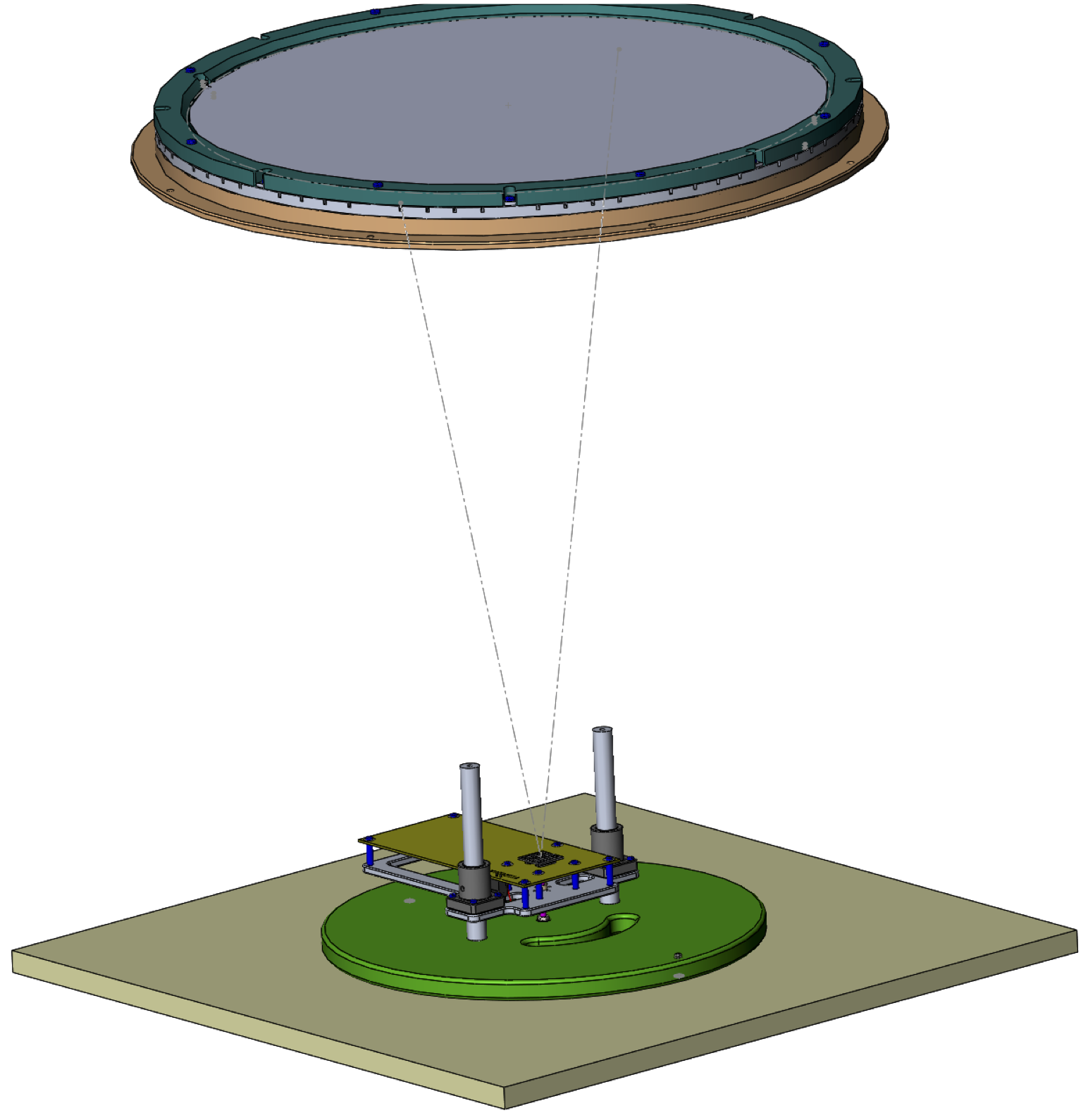} 
\caption{Side View}
\end{subfigure}
\begin{subfigure}{0.49\textwidth}
\centering
\includegraphics[scale=0.3]{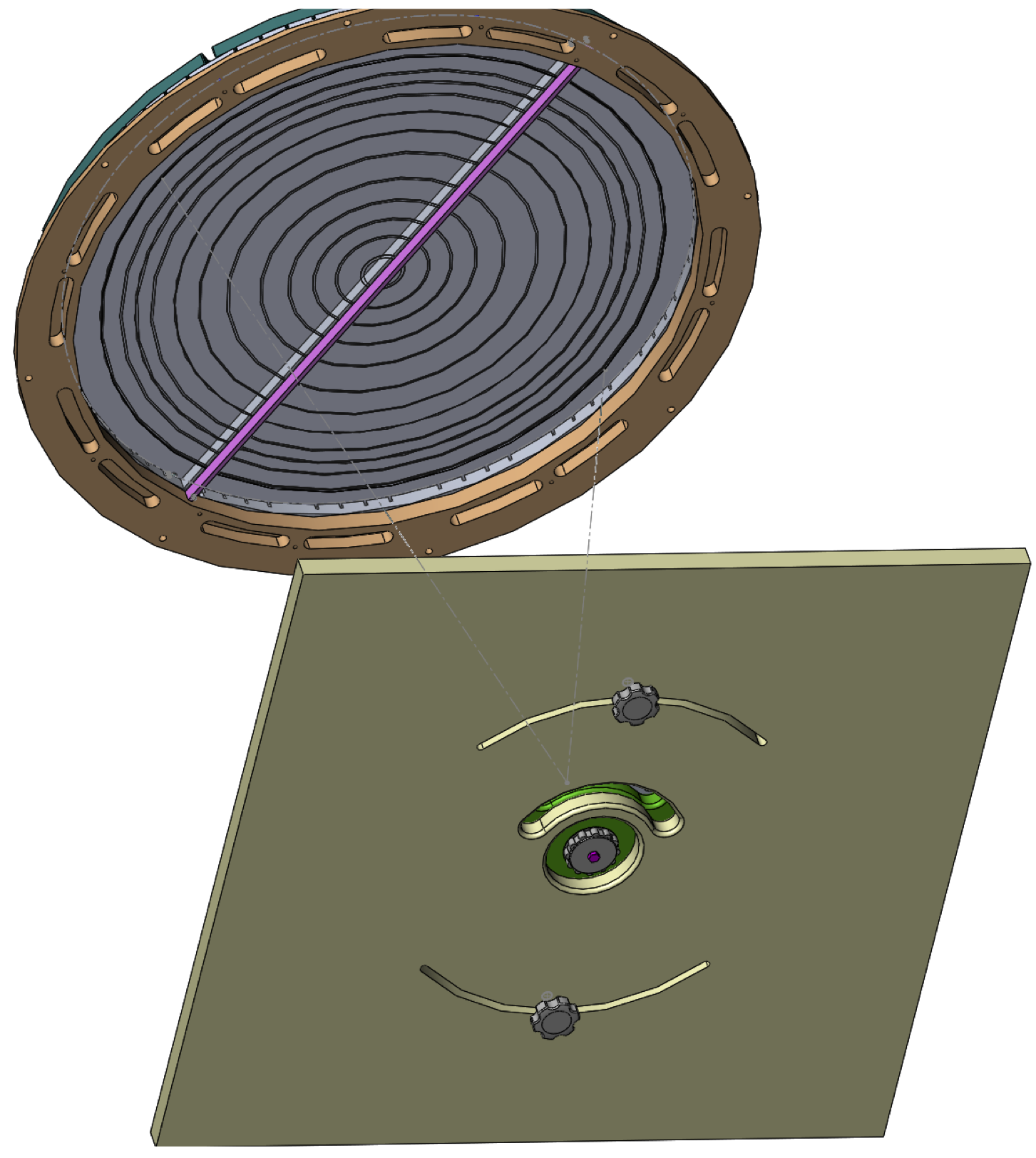}
\caption{Bottom View}
\end{subfigure}
\caption{Prototype module focus stage mock-up. (a): Prototype electronics board shown in yellow with detector array in gray at focus position between aluminum support rods. Stiffness is increased through mounting to the aluminum plate below (gray). Support rods connect to rotating aluminum plate (green) for PA adjustments. A lead screw passes through the rotating plate to connect the aluminum stiffening board to a focus knob on back of the rotating plate. (b): Bottom view of the same focus stage showing the focus knob (center) and paths for rotating the focus stage. The two outer knobs connecting to opposite sides of rotating plate will be loosened to move along the cut-out in wood box. When the desired pointing is reached, both knobs will be tightened to fix the orientation of the detector and focus stage. An opening in both the rotating plate and wood box allows cables to pass through to the electronics board.}
\label{fig:proto_focus}
\end{figure}

For the Prototype module a Hamamatsu SiPM detector will be mounted with the quadrant board\cite{Wright2018} on an aluminum supporting plate attached to an adjustable focus stage at the back of the module via two support rods on which it is free to move vertically. The rods will connect to a rotating back plate to adjust PA, and a lead screw will connect the support plate at a third position and pass to a manual focus knob on the outside of the module box. See Figure \ref{fig:proto_focus} for sketches of this focus stage.

For the final module design, the focus stage will keep the same concept as the Prototype module, but modified for a larger detector and electronics configuration as well as motor-driven focus adjustments. The larger board will still be supported by an aluminum stiffening plate, but now with the detector at the center of a square configuration. This plate will be supported on two sides by four aluminum rods (two support rods on two opposing sides of the plate) with two stepper motors driving lead screws to adjust the focus position. This focus stage may be mounted on a rotating plate like the Prototype module to change the PA, or the focus stage may be fixed directly to the back of the module frame. In the latter case the entire module will be rotated to the desired PA when it is mounted to the geodesic dome. 
\subsection{Baffling}\label{sec:baffling}
Typically, an enclosed dome with only a small opening to the sky helps to mitigate the problem of scattered light reaching the detector, but with our design of many modules pointing in different directions it is not possible to have an enclosure blocking oblique light. For this reason we must baffle both the modules and detectors from scattered light in both optical and near-infrared wavelengths.

For the Beta-1 module, the electronics are housed inside the module frame, requiring the baffling to be easily removable in order to access the electronics. Due to the variety of tests performed with this module these components need to be changed frequently. To accomplish this we designed a flexible neoprene shroud to wrap around the entire module and close with snaps. Snaps were attached to the hexagonal lens mount and the plate at the back of the frame. The shroud is attached to these and wrapped around the module like a jacket with a seam of snaps to close it. Transmission of the neoprene was tested with an infrared 1.5\,$\mu$m, 2\,mW laser and the Ninox 640 InGaAs camera. Light could not be detected passing through 1mm thick neoprene, but 1.5\,mm thick neoprene was used to increase durability for repeated removal.

For the Prototype module, the wood box of the frame will double as baffling so no additional measures will be needed. Both the lens mount, aluminum struts, and electronics/focus stage will all be housed inside the box.  For the final module design however, baffling will be required as the wood box mounting mechanism will not be included.
 
In addition, the focus stage and electronics being located behind the back plate of the frame instead of in front like in the Beta-1 module provide an additional complication. This leads to the overall length of the module components changing slightly with focus position. In addition, the tight spacing of modules in the geodesic dome makes the ability to access the electronics without modifying the entire module advantageous. For these reasons, two separate baffling components may be used for the final module: one to protect the space between the lens mount and back plate from scattered light, and another to protect the electronics and focus stage.

Black anodized aluminum panels will likely be used on the sides of the module frame as shown in Figure \ref{fig:final_module} to baffle the majority of the module. An aluminum box will also likely be used to enclose the electronics and focus stage with the possible addition of a fan depending on the heat dissipation of the final board design. The separation of these components will allow for servicing of electronic or opto-mechanical module components without disturbing the other. The use of aluminum instead of neoprene like in the Beta-1 module will provide improved durability for the lifetime of the PANOSETI instrument.

\begin{table}[h!]
\small
\begin{center}
\caption{Lens Mechanical Properties}
\begin{tabular}{|l|c|c|}
\hline
Property & Edmund Optics 46-392 & Orafol SC214\\
\hline
\hline
Diameter [mm] & 470 & 461 \\
\hline
Focal Length [mm] & 457 & 608 \\
\hline
F-number & 0.97 & 1.32 \\
\hline
Thickness [mm] & 3 & 1.8 \\
\hline
Groove Facet Spacing [mm] & 0.17 & 0.51 \\
\hline
Diameter Thermal Expansion (up to 40$^\circ$C) [mm] & 0.55 & 0.55 \\
\hline
Diameter Thermal Contraction (down to $-$20$^\circ$C) [mm] & -1.4 & -1.4 \\
\hline
Deflection at Center (Gravity) [mm] & 2.3 & 6.2 \\
\hline
Deflection at Center, with Stiffening Beam (Gravity) [mm] & 0.1 & 0.09 \\
\hline
Deflection at Center ($\rm80 \ km \ h^{-1}$ Wind) [cm] & 25 & 11 \\
\hline
Deflection at Center, with Stiffening Beam ($\rm80 \ km \ h^{-1}$ Wind) [mm] & 1.3 & 1.3 \\
\hline
\end{tabular}
\label{tbl:properties}
\end{center}
\end{table}

\section{Summary}\label{sec:summary}

The Beta-1 module was designed as a proof of concept for employing a single Fresnel lens in an optical telescope and for initial on-sky measurements at Mt.~Laguna Observatory. An Edmund Optics 46-392 lens was fixed in a hexagonal mount with silicone rubber in oversized slots allowing for thermal expansion while fixing the location of the lens. A protective layer of acrylic was mounted on the outside of this frame to prevent damage to the Fresnel lens. Lightweight, hollow steel rods connect the hexagonal lens mount to an aluminum plate at the back of the module on which the camera or detector can be attached for testing. To prevent scattered light from affecting measurements, a neoprene baffling is wrapped around the module and secured with snaps to be easily removed. This lightweight module can be mounted to both an optical bench for testing or to the wood base for measurements on-sky.

The Prototype module improves on and refines the design of the Beta-1 module for use at a fixed location. Aluminum rings with larger diameters than the Orafol SC214 Fresnel lens make up the lens mount. The larger diameter allows for free expansion of the lens and protective acrylic plate with changes in temperature. Set screws fix the positions of the Fresnel lens and protective acrylic while stiffening beams prevent unacceptable deflection due to gravity or wind. A manually adjustable focus stage will be included at the back of the module mounted on a rotating back plate. The module will be housed in a wood box for baffling and attaching to a Dobsonian mount for observations. Four of these modules will be used for the prototype phase with two located at each observatory for testing coincidence detections at one location and between the two future domes.

The final module design will make use of the lens mount in the Prototype module with minor adjustments determined as necessary. Aluminum struts will connect the lens mount to an aluminum plate at the back of the module with an opening for light to pass through to the detector. A focus stage modified from the Prototype module to accommodate a larger detector array will be mounted behind this plate and adjusted with two stepper motors. Baffling of the module frame and electronics will be accomplished with two separate components. Some one-hundred of these modules will be incorporated in each geodesic dome--one at Mt.~Laguna and one at Lick Observatory--monitoring for fast transient signals in the entire observable sky at all observable times.

\acknowledgments      
 
The PANOSETI research and instrumentation program is made possible by the enthusiastic support and interest by Franklin Antonio. We thank the Bloomfield Family Foundation for supporting SETI research at UC San Diego in the CASS Optical and Infrared Laboratory. Harvard SETI is supported by The Planetary Society. UC Berkeley's SETI efforts involved with PANOSETI are supported by NSF grant 1407804, the Breakthrough Prize Foundation, and the Marilyn and Watson Alberts SETI Chair fund. Lastly, we would like to thank the staff at Mt.~Laguna and Lick Observatories for their help with equipment testing.

\bibliography{report}
\bibliographystyle{spiebib}

\end{document}